\documentclass[10pt, conference]{IEEEtran}
\IEEEoverridecommandlockouts

\usepackage[T1]{fontenc}
\usepackage[utf8]{inputenc}
\usepackage{amsmath,amssymb,amsfonts}
\usepackage{graphicx}
\usepackage{booktabs}
\usepackage{multirow}
\usepackage{array}
\usepackage{siunitx}
\usepackage{xcolor}
\usepackage{hyperref}
\usepackage{cite}
\usepackage{microtype}
\usepackage{enumitem}
\usepackage{url}
\usepackage{float}
\usepackage{subcaption}
\usepackage{algorithmic}
\usepackage{textcomp}
\usepackage{xcolor}
\usepackage{physics}
\usepackage{bbm}
\usepackage{amsthm}
\usepackage{tikz}
\def\BibTeX{{\rm B\kern-.05em{\sc i\kern-.025em b}\kern-.08em
    T\kern-.1667em\lower.7ex\hbox{E}\kern-.125emX}}

\hypersetup{
    colorlinks=true,
    linkcolor=blue,
    citecolor=blue,
    urlcolor=blue
}

\newcommand{\figdir}{figures}

\newcommand{\safeincludegraphics}[2][]{%
    \IfFileExists{#2}{%
        \includegraphics[#1]{#2}%
    }{%
        \fbox{%
            \parbox[c][4.2cm][c]{0.92\linewidth}{%
                \centering
                Missing figure file:\\[2mm]
                \texttt{\detokenize{#2}}%
            }%
        }%
    }%
}

\makeatletter
\def\bstctlcite#1{\@bsphack
  \@for\@citeb:=#1\do{%
    \edef\@citeb{\expandafter\@firstofone\@citeb}%
    \if@filesw\immediate\write\@auxout{\string\citation{\@citeb}}\fi}%
  \@esphack}
\makeatother
\begin{document}
\bstctlcite{IEEEexample:BSTcontrol}

\title{End-to-End Quantum Semantic Communication with Variational Quantum Neural Networks \\

\thanks{The authors M. Krichen and S. Cherkaoui acknowledge support from the Natural Sciences and Engineering Research Council of Canada (NSERC) and Thales Canada Defence \& Security. The authors N. Nunavath, R. Bassoli, and F. H. P. Fitzek acknowledge the financial support of the German Research Foundation (DFG, Deutsche Forschungsgemeinschaft) under Project ID 390696704, Cluster of Excellence Centre for Tactile Internet with Human-in-the-Loop'' (CeTI) of TUD Dresden University of Technology. This work was also partially funded by the joint project 6G-life (Project ID 16KISK001K) and Disruptive Kommunikationsparadigmen für technologische Souveränität -- CommUnity'' (Project ID 16KISS012K). This work was also supported by a Mitacs Globalink Research Award (Application Ref. IT49074; Funding Request Refs. FR161447 and FR161448).}

}

\author{
\IEEEauthorblockN{
Melek Krichen\IEEEauthorrefmark{1}\IEEEauthorrefmark{4},
Nikhitha Nunavath\IEEEauthorrefmark{2}\IEEEauthorrefmark{3},
Riccardo Bassoli\IEEEauthorrefmark{2}\IEEEauthorrefmark{3},
Soumaya Cherkaoui\IEEEauthorrefmark{1}\IEEEauthorrefmark{4},
Frank H. P. Fitzek\IEEEauthorrefmark{2}\IEEEauthorrefmark{3}
}
\IEEEauthorblockA{
\IEEEauthorrefmark{1} Department of Computer and Software Engineering, Polytechnique Montr\'eal, Montr\'eal, Canada\\
\IEEEauthorrefmark{4} LINCS Lab\\
\IEEEauthorrefmark{2} Deutsche Telekom Chair of Communication Networks,
Technische Universit\"at Dresden, Germany\\
\IEEEauthorrefmark{3} Centre for Tactile Internet with Human-in-the-Loop (CeTI),
Dresden, Germany\\
Emails:
\texttt{\{melek.krichen,soumaya.cherkaoui\}@polymtl.ca}\\
\texttt{\{nikhitha.nunavath,riccardo.bassoli,frank.fitzek\}@tu-dresden.de}
}
}

\maketitle

\begin{abstract}
Quantum-enabled learning is increasingly being explored for future communication and networking applications, including distributed sensing, Internet of Things (IoT), and distributed quantum computing. However, existing approaches often face challenges in scalable inference and efficient information processing. To address these limitations, this paper integrates quantum machine learning (QML) with quantum semantic communication (QSemCom) in an end-to-end learning framework. A classical dataset is first mapped to a low-dimensional feature representation and encoded by a variational quantum transmitter that learns task-relevant semantic features. The resulting features are transmitted through a quantum channel and processed by a receiver to perform a downstream classification task. The proposed framework is evaluated using the MNIST dataset and variational quantum neural networks (QNNs) under ideal and noisy quantum-channel conditions. First, a baseline model is trained over an ideal channel and evaluated under increasing depolarizing noise without retraining. A second set of experiments introduces a trainable receiver QNN and jointly optimizes the transmitter and receiver through end-to-end (E2E) training. The results demonstrate that jointly trained quantum semantic transceivers can adapt the transmitted representation to channel impairments and preserve task-relevant information, highlighting the potential of receiver-aware QML for robust inference in quantum semantic communication.

\end{abstract}


\section{Introduction}
The evolution from fifth-generation (5G) toward sixth-generation (6G) networks is expected to support a new generation of intelligent and immersive services, including extended reality (XR), digital twins, autonomous systems, integrated sensing, and massive Internet of Everything (IoE) connectivity  \cite{strinati20216gnetworksshannonsemantic}, \cite{article}. These applications impose increasingly stringent requirements on data rate, latency, reliability, energy efficiency, sensing, computing, and intelligence. Conventional communication systems, however, largely follow Shannon's paradigm of reliably reproducing transmitted information at the receiver, without explicitly considering its meaning or relevance to a downstream task \cite{10.1109/TIT.1979.1055985}. As future networks become increasingly driven by artificial intelligence (AI) and machine learning (ML), transmitting every bit with equal importance may therefore be inefficient when only task-relevant information is required.

Semantic communication (SemCom) addresses this gap by shifting the communication objective from reliable bit transmission toward the accurate transmission of meaning, i.e., the semantic content relevant to the receiver, thereby potentially reducing communication overhead \cite{article, strinati20216gnetworksshannonsemantic}. First introduced theoretically in \cite{6004632}, SemCom has since developed into an active research area, with recent work formalizing the foundations of semantic information theory and extending Shannon's framework toward meaning-aware communication \cite{shao2025theorysemanticcommunication, 9723337}. Machine learning has become a major enabling approach for extracting compact semantic representations and optimizing semantic encoding and decoding. In particular, knowledge graphs (KGs) \cite{hello2024semanticcommunicationenhancedknowledge, 11101041, e24060846}, graph Neural Networks (GNNs) \cite{zheng2024genetgraphneuralnetworkbased, ir.2025.41} and large language models (LLMs) \cite{jin2024largelanguagemodelsgraphs, Wang_2025} have been explored for representing entities and their semantic relationships, with the aim of improving semantic compression and robustness against communication noise \cite{9679803}.

In parallel, quantum computing and quantum communication have emerged a rapidly growing fields that exploit quantum mechanical properties such as superposition, entanglement and quantum parallelism \cite{unknown, wang2025quantumsemanticcommunicationshannonwyner,10288526} for information processing and communication. These advances have motivated the integration of quantum technologies with semantic communication. Prior work has proposed the conceptual frameworks for QSemComs based on quantum embedding, quantum feature extraction, quantum machine learning, and semantic representations \cite{10251843, chehimi2024quantumsemanticcommunicationsresourceefficient,article}. Other studies have investigated graph-based approaches involving knowledge graphs, binary quantum encoding and quantum channel coding \cite{10694659,11101041, unknown}, as well as more recent directions  in resource-efficient quantum semantic networking and secure semantic communications for future 6G networks \cite{10968853,wang2025quantumsemanticcommunicationshannonwyner,11162234}.

A relevant application scenario for QSemCom is a distributed quantum computing network in which multiple quantum processing units (QPUs) are located at different nodes and exchange quantum information over quantum channels, such as optical-fiber links \cite{11142688}. Although such networks may contain multiple interconnected nodes, this work focuses on a two-node setting in which one QPU operates as the transmitter and another as the receiver, connected through a quantum channel.

Despite these advances, classical semantic communication frameworks primarily rely on deep neural networks, transformer-based language models, GNNs, and KGs for semantic extraction, encoding, and transmission over classical channels. Recent QSemCom frameworks, in contrast, have largely relied on predefined mechanisms, such as deterministically prepared quantum states, binary quantum encoding, quantum clustering, and quantum teleportation, where the quantum communication process itself is not jointly optimized for the underlying semantic learning task. Moreover, existing evaluations mainly focus on semantic fidelity, reconstruction quality, communication efficiency, or resource consumption, leaving the E2E trainability of VQNNs directly under noisy quantum channel conditions largely unexplored.

To address this gap, we propose an E2E QSemCom framework based on trainable VQNNs operating under noisy quantum channels. A classical semantic encoder first compresses the input into a low-dimensional latent representation, which is then mapped to a variational quantum transmitter for semantic encoding and transmission. After propagation through the noisy quantum channel, a trainable variational quantum receiver processes the received information before performing the downstream classification task. We investigate ideal, bit-flip, depolarizing, and amplitude-damping quantum channels under two experimental settings. First, a baseline transmitter (TX-QNN) is trained over an ideal channel and evaluated under increasing channel noise without retraining. Second, an E2E recovery strategy jointly optimizes the transmitter and receiver (TX-QNN and RX-QNN) under noisy channel conditions, allowing the learned quantum transceiver to compensate for channel-induced degradation and preserve task-relevant information.

The main contributions of this paper are summarized as follows:

\begin{itemize}
\item We propose an QSemCom architecture integrating a classical semantic encoder with trainable VQNN at the transmitter side.

\item We evaluate the framework under noisy channels and demonstrate high semantic classification performance under weak to moderate quantum noise.

\item We introduce an E2E recovery strategy using a trainable receiver-side quantum neural network to compensate for channel-induced noise.
\end{itemize}

\section{Proposed Framework}
\subsection{System Overview}
\begin{figure*}[h]
    \centering
    \includegraphics[width=\textwidth]{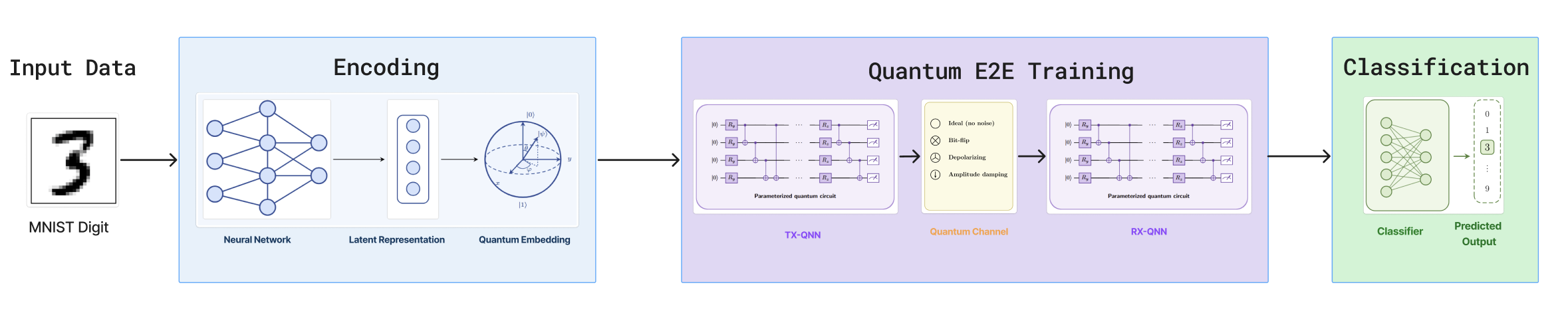}
    \caption{Proposed E2E QSemCom framework. A classical semantic encoder first compresses the input into a low-dimensional latent representation, which is transformed into quantum states through angle encoding before being processed by the transmitter-side QNN. The encoded quantum state is transmitted through a quantum channel and recovered by a receiver-side VQNN. Finally, a classifier predicts the semantic class from the recovered semantic representation.}
    \label{fig:pipeline}
\end{figure*}

The framework shown in Fig.~\ref{fig:pipeline} consists of three steps. The first step is a classical encoding phase, where the input data is compressed into a latent representation and encoded into quantum states. The second step is an end-to-end (E2E) semantic transmission phase, where the encoded quantum states are processed by a TX-QNN, propagated through a quantum channel, and recovered by an RX-QNN trained to compensate for channel-induced distortions while preserving task-relevant semantic information. The final step is the classification of the recovered quantum features.

\subsection{Dataset}

The proposed framework is evaluated on the MNIST dataset \cite{6296535}, which consists of grayscale images of digits from 0 to 9. Each image has a resolution of $28\times28$ pixels and belongs to one of ten classes.

\subsection{Classical Encoder}

The classical encoder compresses the high-dimensional input data into a compact latent representation that preserves semantic information. Given an input image $\mathbf{X}\in\mathbb{R}^{H\times W}$, the image is flattened into a one-dimensional vector $\mathbf{x}\in\mathbb{R}^{HW}$ and processed by the encoder to generate a low-dimensional latent representation $\mathbf{z}=f_{\mathrm{enc}}(\mathbf{x})$, where $\mathbf{z}\in\mathbb{R}^{d}$ and $d\ll HW$. The latent vector carries the semantic information derived from the input image. The encoder is a fully connected neural network independently trained prior to the quantum communication experiments.

\subsection{Quantum Embedding}

The latent representation is encoded into a quantum state through angle encoding. The value of each element in the latent space corresponds to the rotation angle of a one-qubit gate operation, which creates the quantum state for transmission.

\begin{equation}
 z \longmapsto \lvert\psi_0\rangle = R_k(z)\lvert 0\rangle 
\end{equation}

where $R_k(\cdot)$ represents a rotation around the axis $k\in\{x,y,z\}$ on the Bloch sphere.

\subsection{Variational Quantum Transmitter (TX-QNN)}

Following quantum state preparation, the encoded quantum state is processed by the TX-QNN. The transmitter consists of parameterized single-qubit rotation gates and entangling operations, resulting in the unitary transformation:

\begin{equation}
|\psi_{\mathrm{tx}}\rangle
=
U_{\mathrm{TX}}(\boldsymbol{\theta}_{\mathrm{tx}})
\lvert\psi_0\rangle
\end{equation}

\noindent where $\boldsymbol{\theta}_{\mathrm{tx}}$ denotes the TX-QNN parameters. The transmitter transforms the angle-encoded quantum state into a semantic quantum representation in the Hilbert space by learning semantic features through parameterized quantum
operations.

\subsection{Quantum Channel}

The quantum state is transmitted through a quantum channel and can be affected by environmental noise \cite{sakai2026quantumhardwarenoiselearning, li2025languagemodellargetexttransmission, 10251843}. The transmitted quantum state is represented by a density matrix operator $\rho_{\mathrm{tx}} = |\psi_{\mathrm{tx}}\rangle \langle \psi_{\mathrm{tx}}|$, and its evolution is defined as: 

\begin{equation}
\rho_{\mathrm{rx}}
=
\mathcal{N}(\rho_{\mathrm{tx}})
=
\sum_{i=0,1}
K_i
\rho_{\mathrm{tx}}
K_i^\dagger
\end{equation}

where $\rho_{\mathrm{tx}}$ and $\rho_{\mathrm{rx}}$ are the transmitted and received density matrices respectively, $\mathcal{N}(\cdot)$ is the quantum channel and $K_i$ is the Kraus operator satisfying $\sum_i K_i^\dagger K_i = I$ for all $i \in \{0,1\}$ \cite{sakai2026quantumhardwarenoiselearning, li2025languagemodellargetexttransmission}. For our numerical simulations, we considered four quantum channel models \cite{10251843, li2025languagemodellargetexttransmission, 11101041, sakai2026quantumhardwarenoiselearning} defined as:

\subsubsection{Ideal Channel}

Represents noise-free communication channel where: 

\begin{equation}
\rho_{\mathrm{rx}}
=
\rho_{\mathrm{tx}}
\end{equation}

\subsubsection{Bit-Flip Channel}
The random bit-flip channel models errors with error probability $p$ as 

\begin{equation}
K_0
=
\sqrt{1-p}\,I,
\qquad
K_1
=
\sqrt{p}\,X
\end{equation}

\subsubsection{Depolarizing Channel}

The depolarizing channel models isotropic noise through Pauli errors, and is defined as: 

\begin{equation}
\mathcal{N}(\rho)
=
(1-p)\rho
+
\frac{p}{3}
\left(
X\rho X
+
Y\rho Y
+
Z\rho Z
\right)
\label{eq:depo_channel}
\end{equation}

\subsubsection{Amplitude-Damping Channel}

This channel models the amplitude damping probability of quantum states undergoing energy relaxation from excited state $|1\rangle$ to ground state $|0\rangle$:

\begin{equation*}
K_0=
\begin{bmatrix}
1&0\\
0&\sqrt{1-p}
\end{bmatrix},
\qquad
K_1=
\begin{bmatrix}
0&\sqrt{p}\\
0&0
\end{bmatrix}
\end{equation*}

where $p$ denotes the damping probability parameter.

\subsection{Variational Quantum Receiver (RX-QNN)}

The RX-QNN is introduced to recover task-relevant semantic information that was degraded during transmission. Given the received quantum state $\rho_{\mathrm{rx}}$, the receiver applies a trainable variational circuit to compensate for channel-induced distortions: 

\begin{equation}
\rho_{\mathrm{out}}
=
U_{\mathrm{RX}}
(\boldsymbol{\theta}_{\mathrm{rx}})
\,
\rho_{\mathrm{rx}}
\,
U_{\mathrm{RX}}^\dagger
(\boldsymbol{\theta}_{\mathrm{rx}})
\end{equation}

where $\boldsymbol{\theta}_{\mathrm{rx}}$ denotes the trainable receiver parameters.

\subsection{Classifier}

After receiver processing, expectation values of the measured observables are collected to form a classical feature vector:

\begin{equation}
\mathbf{h}
=
\left[
\langle Z_1\rangle,
\ldots,
\langle Z_n\rangle
\right]
\end{equation}

These recovered semantic features are provided to a fully connected classifier that predicts the class:

\begin{equation}
\hat{\mathbf{y}}
=
f_{\mathrm{cls}}(\mathbf{h})
\end{equation}

\subsection{Learning Objective}

The entire framework is optimized using supervised E2E learning by minimizing the categorical cross-entropy loss:

\begin{equation}
\mathcal{L}
=
-
\sum_{i=1}^{C}
y_i
\log(\hat{y}_i)
\end{equation}

where $C$ denotes the number of classes.


\section{Experimental Setup}

The proposed QSemCom framework is implemented using PyTorch \cite{NEURIPS2019_bdbca288} for the classical neural network components and PennyLane \cite{bergholm2018pennylane} for quantum circuits simulations.

\subsection{Classical Encoder}

The classical encoder is a fully connected neural network that compresses the flattened image into a low-dimensional latent representation. A Tanh activation constrains the latent values to $[-1,1]$. The architecture is summarized in Table~\ref{tab:semantic_encoder}.

\begin{table}[h]
\centering
\caption{Architecture of the classical semantic encoder.}
\label{tab:semantic_encoder}
\begin{tabular}{ll}
\toprule
\textbf{Layer} & \textbf{Configuration} \\
\midrule
Input & $28\times28$ grayscale image \\
Flatten & $28\times28 \rightarrow 784$ \\
Fully connected 1 & $784 \rightarrow 256$ \\
Activation & ReLU \\
Fully connected 2 & $256 \rightarrow 128$ \\
Activation & ReLU \\
Fully connected 3 & $128 \rightarrow d$ \\
Output activation & Tanh \\
Output & Latent vector of dimension $d$ \\
\bottomrule
\end{tabular}
\end{table}

\subsection{Quantum Circuits Configuration}

\begin{figure}[h]
    \centering
    \includegraphics[width=1\linewidth]{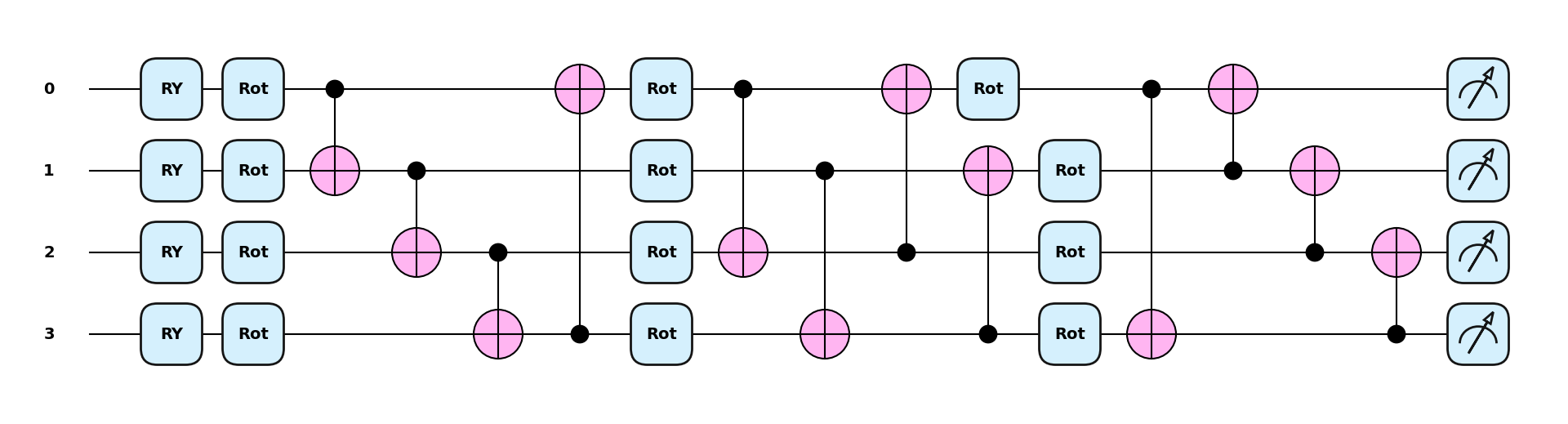}
    \caption{Transmitter Variational Quantum Circuit (TX-QNN)}
    \label{fig:tx}
\end{figure}

The framework contains two QNNs. The TX-QNN initializes the quantum state through angle encoding with $R_Y$ rotations and applies variational layers using the \texttt{StronglyEntanglingLayers} template of PennyLane. Data re-uploading is supported when the latent dimension exceeds the number of qubits. The transmitter and receiver operate on four qubits with three and two variational layers, respectively. The transmitter processes the encoded state before transmission, while the receiver operates directly on the distorted state. The TX-QNN is illustrated in Fig.~\ref{fig:tx}.

The ideal channel is simulated using \texttt{lightning.qubit}, while noisy channels use the density matrix based \texttt{default.mixed} backend. The complete configuration is summarized in Table~\ref{tab:qnn_architecture}.

\begin{table}[h]
\centering
\caption{Quantum circuit architecture.}
\label{tab:qnn_architecture}
\begin{tabular}{ll}
\toprule
\textbf{Component} & \textbf{Configuration} \\
\midrule
Quantum encoding & Angle encoding ($R_Y$ rotations) \\
Latent space dimension & 4 \\
Number of qubits & 4 \\
Transmitter layers & 3 \\
Receiver layers & 2 \\
TX parameters & 36 \\
RX parameters & 24 \\
TX variational block & StronglyEntanglingLayers \\
RX variational block & StronglyEntanglingLayers \\
Entangling gates & CNOT \\
Measurement & Pauli-$Z$ expectation values \\
Quantum simulator & \texttt{lightning.qubit} / \texttt{default.mixed} \\
\bottomrule
\end{tabular}
\end{table}

\subsection{Training}

Two training strategies are investigated. First, the classical encoder, TX-QNN, and classifier are trained over an ideal channel and then evaluated under increasing noise without retraining. Second, the TX-QNN, RX-QNN, and classifier are trained jointly. The best model is selected based on the test F1-score. The training configurations are summarized in Table~\ref{tab:training}.

\begin{table}[h]
\centering
\caption{Training configuration.}
\label{tab:training}
\begin{tabular}{ll}
\toprule
\textbf{Parameter} & \textbf{Value} \\
\midrule
\multicolumn{2}{c}{\textbf{Baseline TX Training}}\\
\midrule
Optimizer & AdamW \\
Batch size & 64 \\
Epochs & 100 \\
Learning rate & $10^{-4}$ \\
Quantum channel & Perfect \\
Best model criterion & F1-score \\
Weight decay & $10^{-4}$ \\
Label smoothing & 0.1 \\
\midrule
\multicolumn{2}{c}{\textbf{E2E (TX-RX) Training}}\\
\midrule
Optimizer & AdamW \\
Batch size & 64 \\
Epochs & 30 \\
Learning rate & $10^{-4}$ \\
Quantum channel & Depolarizing \\
RX variational layers & 2 \\
Best model criterion & F1-score \\
Weight decay & $10^{-4}$ \\
Label smoothing & 0.1 \\
\bottomrule
\end{tabular}
\end{table}

\subsection{Classifier}

The Pauli-$Z$ expectation values measured on the four qubits form a four-dimensional quantum feature vector processed by a fully connected classifier. The classifier is jointly optimized with the trainable components using categorical cross-entropy loss. The classifier architecture is summarized in Table~\ref{tab:classifier}.

\begin{table}[h]
\centering
\caption{Architecture of the semantic classifier.}
\label{tab:classifier}
\begin{tabular}{ll}
\toprule
\textbf{Layer} & \textbf{Configuration} \\
\midrule
Input & 4 quantum features \\
Fully connected 1 & $4 \rightarrow 64$ \\
Activation & ReLU \\
Dropout & 0.35 \\
Fully connected 2 & $64 \rightarrow 32$ \\
Activation & ReLU \\
Dropout & 0.25 \\
Output layer & $32 \rightarrow 10$ \\
Output & Class logits \\
\bottomrule
\end{tabular}
\end{table}

\subsection{Evaluation Metrics}

\paragraph{Accuracy}

Accuracy measures the proportion of correctly classified samples over the entire test set, providing an overall measure of the model's classification performance.

\paragraph{F1-score}

The F1-score combines precision and recall into a single metric, accounting for both false positives and false negatives. In this work, the macro F1-score is used by computing the F1-score independently for each class and averaging across all classes, giving equal importance to each class regardless of its frequency.


\paragraph{Signal-to-Noise Ratio}

For the depolarizing channel \cite{11101041, 11162234, hello2024semanticcommunicationenhancedknowledge}, the state remains unchanged with probability $(1-p)$, while each Pauli error occurs with probability $p/3$ \cite{11101041}. The SNR is derived from Equation \ref{eq:depo_channel} as \cite{11101041}:

\begin{equation}
\mathrm{SNR}_{\mathrm{dep}}
=
\frac{1-p}{p/3}
=
\frac{3(1-p)}{p}
\end{equation}

\noindent and reported in decibels as:

\begin{equation}\label{Eqn-SNR-depolarising}
\mathrm{SNR}_{\mathrm{dep,dB}}
=
10\log_{10}
\left(
\frac{3(1-p)}{p}
\right)
\end{equation}

\noindent Accuracy and F1-score are evaluated as functions of the SNR.

\section{Simulation Results and Analysis}

\subsection{Perfect-Channel Training and Robustness Analysis}

The first experiment evaluates the proposed QSemCom framework under ideal quantum communication conditions. In this setting, the TX-QNN is trained over a noise-free quantum channel to learn task-relevant semantic representations and establish a pretrained transmitter for the robustness analysis.

\begin{figure}[h]
\centering
\safeincludegraphics[scale=0.35]{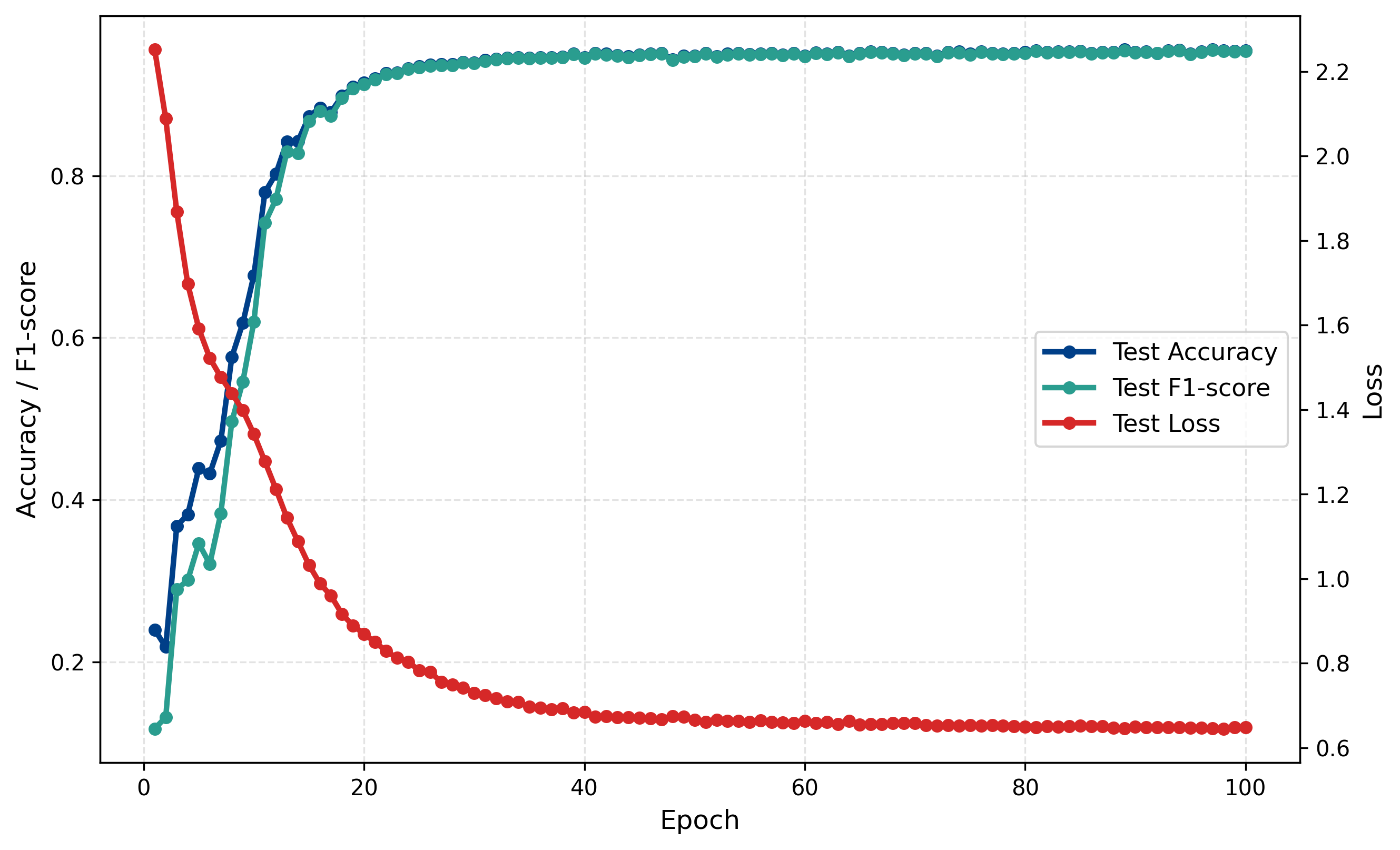}
\caption{ \centering Test loss, accuracy, and F1-score for TX-QNN under the perfect channel.}
\label{fig:perfect_metrics}
\end{figure}

Figure~\ref{fig:perfect_metrics} shows the evolution of the test loss, accuracy, and F1-score. The test loss decreases rapidly during the first epochs before gradually converging, while accuracy and F1-score increase and stabilize after approximately 50 epochs.

\begin{figure}[h]
\centering
\safeincludegraphics[scale=0.35]{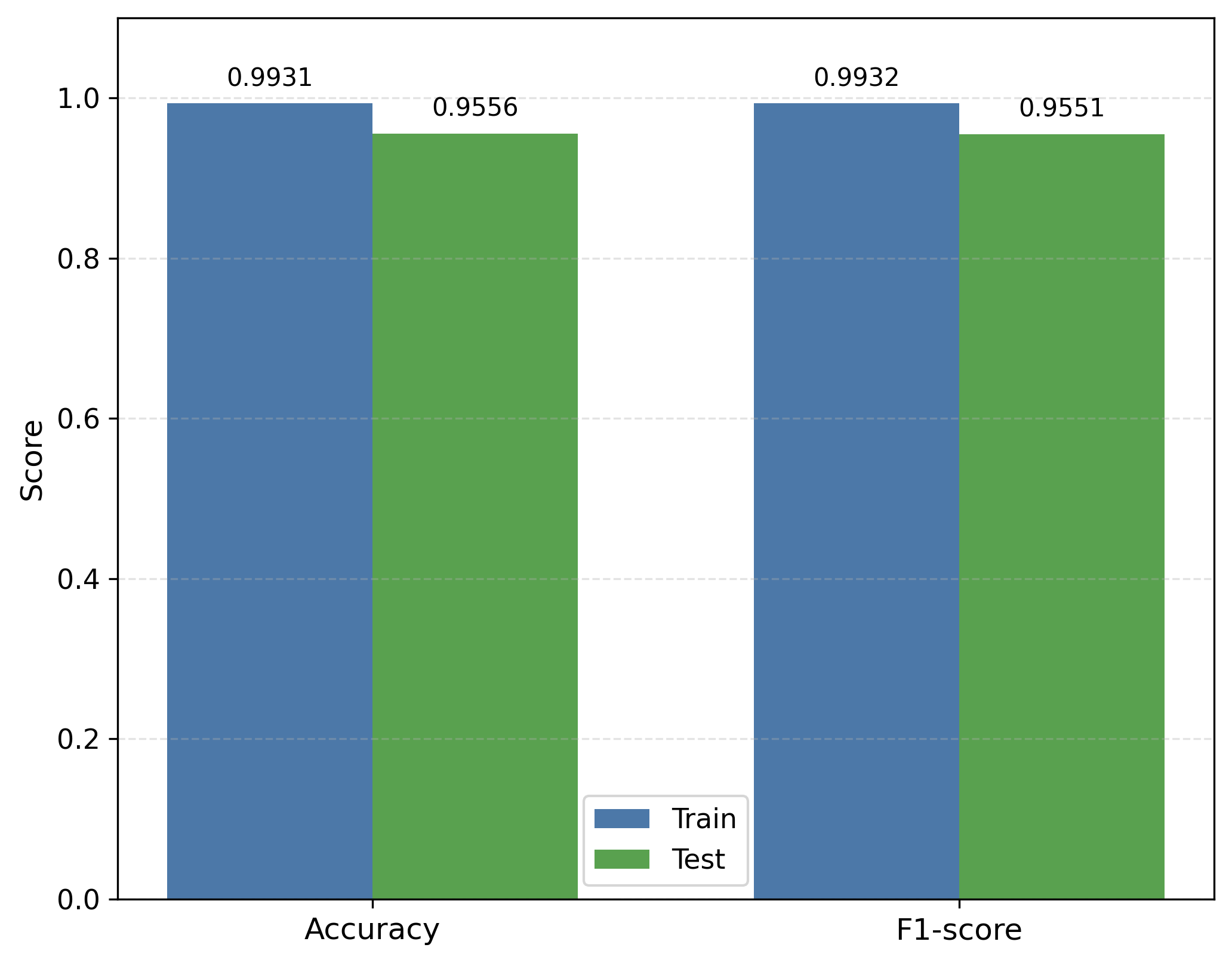}
\caption{Best accuracy and F1-score for TX-QNN.}
\label{fig:bestmetrics}
\end{figure}

Figure~\ref{fig:bestmetrics} provides an overview of the best results obtained during the training. The model attains 0.9556 and 0.9551 of test accuracy and test F1-score, respectively. In an ideal quantum communication scenario, this means that the model is capable of learning quantum semantics in order to keep the important information needed for classification.

\begin{figure}[h]
\centering
\safeincludegraphics[scale=0.35]{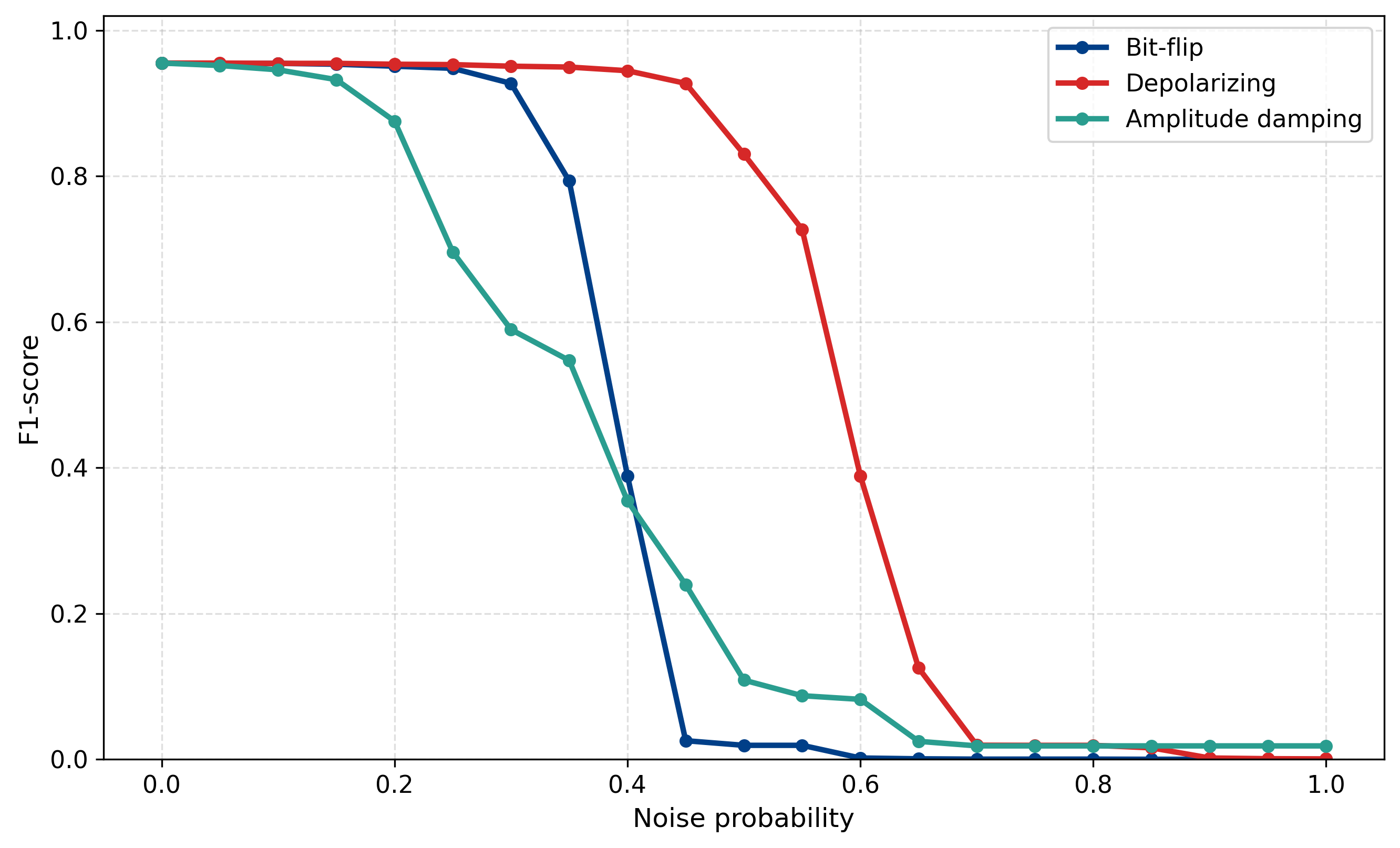}
\caption{F1-score of TX-QNN under noisy channels.}
\label{fig:noisesweepf1}
\end{figure}

We evaluate the pretrained transmitter under bit-flip, depolarizing and amplitude-damping channels. Figure~\ref{fig:noisesweepf1} shows the F1-score as a function of the channel noise probability, accuracy follows the same trend. The depolarizing channel maintains both metrics above $0.9$ until $p=0.45$ while the bit-flip noise shows an earlier decrease at $p=0.25$ and the amplitude damping shows a slower progressive decay. We argue that the differences are due to the influence of each noise process on the learned quantum representation. Bit-flip noise applies Pauli-$X$ errors, on the other hand, amplitude damping drives the qubits to the ground state, losing information. Conversely, depolarizing noise spreads errors across the three Pauli directions, allowing the learned representation to remain task-relevant in the presence of moderate noise.

\subsection{End-to-End Training}

The second set of experiments focuses on the suggested RX-QNN and its ability to extract the task-relevant semantic information after its transmission through the noisy depolarizing quantum channel. In contrast to the first baseline experiment where only the TX-QNN is trained, we introduce a trainable RX-QNN that is jointly trained with the TX-QNN. The goal is to learn the quantum transformation at the receiver end that will compensate for the channel noise and preserve semantic information necessary for subsequent classification.

This same procedure is performed for all considered levels of noise probability. For each level of noise, the optimization of the receiver is performed for the specific channel noise probability and compared with the performance of the perfect channel-trained model. To provide a communication-oriented interpretation, the classification performance is shown as a function of SNR for the depolarizing channel as defined in Equation \ref{Eqn-SNR-depolarising}. For the evaluated noise range ((p=0.1,\ldots,0.9)), the corresponding SNR ranges from 14.31 to -4.77 dB.

\begin{figure}[h]
\centering
\safeincludegraphics[scale=0.38]{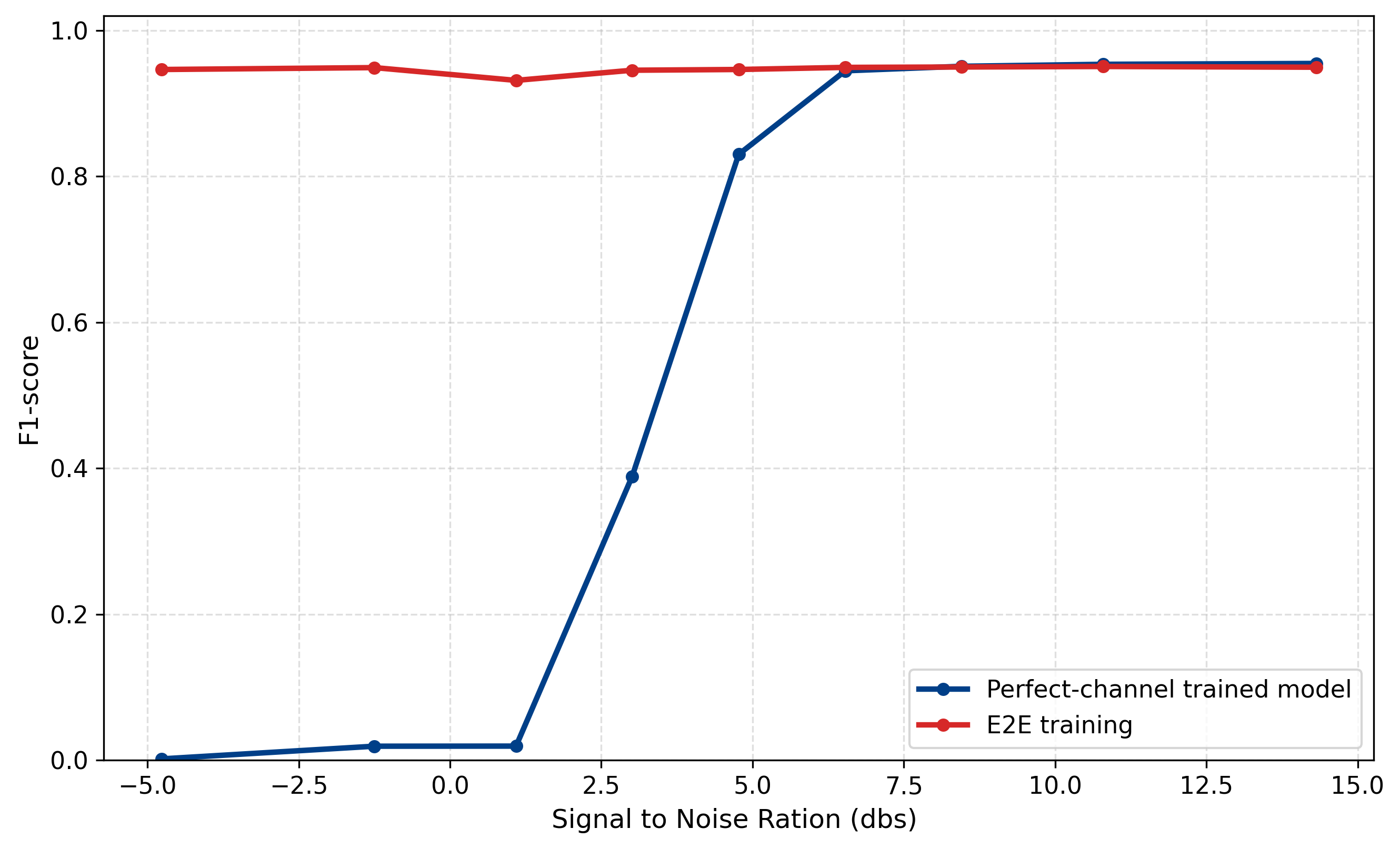}
\caption{F1-score as a function of SNR.}
\label{fig:snrf1}
\end{figure}

Figure~\ref{fig:snrf1} compares the proposed E2E framework with the TX-QNN-only baseline over the same SNR range. At high SNR values, both approaches achieve nearly identical performance. As the SNR decreases, the F1-score of the transmitter-only baseline deteriorates, while the proposed E2E training maintains an F1 close to $95\%$ over the whole SNR range. These results confirm that E2E training improves the robustness of the proposed QSemCom framework under depolarizing noise.

\section{Discussion and Future Directions}

\subsection{Conclusion}

The presented results demonstrate the potential of the end-to-end quantum semantic communication approach utilizing variational quantum neural networks. The TX-QNN trained with perfect channels shows high classification robustness when exposed to weak and medium noise levels. However, the classification accuracy and F1-score become lower with an increase in the noise level. To mitigate this problem, we introduce the RX-QNN, enabling E2E training and significantly improving classification performance under depolarizing noise.

\subsection{Limitations and Future Directions}

Despite the encouraging outcomes, the proposed method is restricted to working on the MNIST dataset, a four-qubit model, quantum channels simulation, and RX-QNNs being trained separately for each depolarizing probability. Additionally, the noise models employed in this work are based on the assumption of independent errors, whereas the quantum hardware used in practice might have much more complex noise behavior. In the future, the proposed methodology should be extended to more complex tasks, including text, speech, and multimodal data, to bigger quantum systems, and to real quantum hardware. One more critical aspect that should be addressed is developing adaptive receivers able to generalize in various channel conditions.

\bibliographystyle{IEEEtran}
\bibliography{references}

\end{document}